\documentclass[twocolumn,english,aps,10pt,superscriptaddress]{revtex4}
\usepackage[T1]{fontenc}
\usepackage{amsmath,graphicx,amssymb,epsfig,babel,dsfont,mathrsfs,color,amsbsy}

\begin{document}

\title{Deep sub-wavelength scale focusing of heat flux radiated by magneto-optical nanoemitters in the presence of an external magnetic-field}

\author{Louis Rihouey}
\affiliation{Laboratoire Charles Fabry, UMR 8501, Institut d'Optique, CNRS, Universit\'{e} Paris-Saclay, 2 Avenue Augustin Fresnel, 91127 Palaiseau Cedex, France}

\author{Philippe Ben-Abdallah}
\email{pba@institutoptique.fr} 
\affiliation{Laboratoire Charles Fabry, UMR 8501, Institut d'Optique, CNRS, Universit\'{e} Paris-Saclay, 2 Avenue Augustin Fresnel, 91127 Palaiseau Cedex, France}

\author{Riccardo Messina}
\email{riccardo.messina@institutoptique.fr}
\affiliation{Laboratoire Charles Fabry, UMR 8501, Institut d'Optique, CNRS, Universit\'{e} Paris-Saclay, 2 Avenue Augustin Fresnel, 91127 Palaiseau Cedex, France}

\date{\today}

\begin{abstract}
We introduce a theoretical framework to describe the heat flux radiated in the near-field regime by a set of magneto-optical thermal nanoemitters close to a substrate in the presence of an external magnetic field. Then, we investigate the particular case of a single emitter and we demonstrate that the external field can induce both an amplification of the heat exchanged between emittter and substrate and a focusing of the Poynting field at the substrate interface at deep sub-wavelength scale. These effects open up promising perspectives for the development of heat-assisted magnetic-recording technology.
\end{abstract}

\maketitle

\section{Introduction}

The near-field scanning thermal microscope~\cite{De Wilde,Achim,Raschke,Hillenbrand,Weng}, a noncontact variant of conventional scanning thermal microscope ~\cite{Williams,Majumdar}, enables local heating at the submicrometric scale by utilizing the tunneling of non-radiative thermal photons (evanescent waves). This near-field technology is promising for nano-photolitography~\cite{Srituravanich04} and for hard-drive writing technology, specifically in heat-assisted magnetic recording (HAMR)~\cite{Challener,Stipe}. In HAMR, a small surface area of a magnetic material is heated to raise its temperature close to the Curie temperature, where its magnetic coercivity is weak. Then by applying a magnetic field a new magnetic state can be recorded inside the material. For high-density magnetic bit storage, the hot spot area should be minimized to approach the superparamagnetic limit, beyond which bits become unstable due to thermal fluctuations. Typically, superparamagnetism in common magnetic materials is observed in domains below 20\,nm in size. However, radiative heat focusing by a conventional scanning probe microscope is constrained by the emission pattern of its tip in the near-field regime. In a recent work~\cite{BenAbdallah19} a theory describing heat flux radiated in the near-field regime by several interacting nanoemitters at different temperatures has been introduced, demonstrating that, in comparison to a single emitter, the thermal energy can be focused and amplified into smaller spots than single emitters paving the way for a multitip near-field scanning thermal microscopy with potential applications in nanoscale thermal management, heat-assisted data recording, nanoscale thermal imaging, heat capacity measurements, and infrared spectroscopy of nanoobjects. The formalism employed in this work falls into the category of near-field radiative heat transfer in dipolar systems, which has attracted a remarkable attention during the last decade~\cite{BenAbdallah11,Messina13,BenAbdallah13b,Tervo19,Luo20,Fang23,Biehs16,Dong18,Messina18,Deshmukh18,Zhang19a,Zhang19b,Ott20,Zhang20,Ott21,Fang22,Asheichyk22,Saaskilahti14,Asheichyk18,Zhang21,Chen22,Zhang23,Asheichyk}. As a matter of fact, the possibility of describing small nanoparticles or objects (such as tips) within the dipolar approximation leads to simpler analytical expressions allowing to unveil interesting two- and $N$-body effects~\cite{Volokitin07,Song15,Joulain05,Cuevas18,Biehs21} and related practical appplications including thermal management~\cite{BenAbdallah2006,Latella21a}, solid-state cooling~\cite{Chen15,Zhu19}, infrared sensing and spectroscopy~\cite{De Wilde06,Jones12}, energy-conversion devices~\cite{DiMatteo01,Narayanaswamy03,Laroche06,Park08,Latella21b} and thermotronics~\cite{BenAbdallah13a,BenAbdallah16PRB,BenAbdallah15,Reddy24}.

In the present work we extend this study to magneto-optical many-body systems that is to the modeling and analysis of heat flux radiated by a set of emitters whose optical properties can be externally manipulated by applying a magnetic field. Many theoretical works have been devoted to date to the control of radiative heat exchanges in systems involving magneto-optical nanoparticles~\cite{Ben-Abdallah2016,Zhu2016,Latella2017,Ekeroth2017,OttEtAl2018,Cuevas,Ott2019a,Ott2019b,Wang20,Messina2021,Lu2022,Ge2023} and to the control of the local density of states of the electromagnetic field~\cite{BenAbdallah2024}. Here we pay a particular attention to the heat flux radiated by these systems in their close environment. As a concrete application, we analyze in detail the heat flux radiated by a single nanoemitter, simulating a heated tip made with a magneto-optical material, above an isotropic substrate and we show that the application of an external field can produce the simultaneous effects of increasing the emitter-to-substrate flux by reducing, at the same time, the spreading of heat flux at the interface of substrate. This focusing effect is investigated with respect to both the size of emitter and the magnitude of applied magnetic field and its origin is explained through a spectral analysis of the heat flux. The paper is structured as follows. In Sec.~\ref{sec:system} we present the physical system and a formalism giving the Poynting vector for a system of $N$ dipoles close to a substrate. The results in the case of a single particle are presented in Sec.~\ref{sec:results}, for different particle radii and as a function of the magnetic field. The last Section presents some conclusive remarks and perspectives.

\section{Physical system and formalism}\label{sec:system}

The system we consider consists of $N$ spherical particles labeled with an index $i=1,2,\dots,N$ having equal radius $R$, coordinates $\mathbf{r}_i$, with $z_i>0$, placed in vacuum and in proximity of a substrate having a frequency-dependent permittivity $\varepsilon_i(\omega)$. The interface between vacuum and substrate coincides with the plane $z=0$. In the following, we describe the formalism to calculate the Poynting vector $\mathbf{S}(\mathbf{r},t)=\mathbf{E}(\mathbf{r},t)\times\mathbf{H}(\mathbf{r},t)$ at an arbitrary point $\mathbf{r}$. This approach, following the one detailed in Ref.~\cite{Messina13}, exploits the fluctuation-dissipation theorem describing the correlation functions of fluctuating dipoles, along with the knowledge of the Green's function of the system, taking into account the presence of a substrate as a boundary condition in the system.

\subsection{Coupled fluctuating dipoles and Green's tensors}

First, we work in the dipolar approximation, in which each individual particle is described in terms an electric dipole. This approach is valid as long as all distances involved (both between particles and between particles and substrate) are large compared to the particle radius $R$. As a rule of thumb, the validity of this approximation is guaranteed when all distances satisfy $d>3R$~\cite{BenAbdallah2008}. Each electric dipoles is decomposed as
\begin{equation}
	\mathbf{p}_i(t) = \mathbf{p}_i^{\text{(fl)}}(t) + \mathbf{p}_i^{\text{(ind)}}(t),
\label{eq:pt}\end{equation}
where the first term accounts for the thermally-fluctuating contribution, whereas the second describes the part induced by the presence of the other dipoles and the substrate. The collection of dipoles generates an electric and magnetic field at any point $\mathbf{r}$, which can be written as 
\begin{equation}\begin{split}
	\mathbf{E}(\mathbf{r},\omega)&=\frac{\omega^2}{\varepsilon_0c^2}\sum_i\mathds{G}^{\text{(EE)}}(\omega,\mathbf{r},\mathbf{r}_i)\mathbf{p}_i,\\
	\mathbf{H}(\mathbf{r},\omega)&=-i\frac{\omega^2}{c}\sum_i\mathds{G}^{\text{(HE)}}(\omega,\mathbf{r},\mathbf{r}_i)\mathbf{p}_i,
\end{split}\label{eq:EHGp}\end{equation}
where the sum extends from 1 to $N$ and we employed a frequency decomposition where each real physical quantity $f(t)$ is written as
\begin{equation}
	f(t)=\int_{-\infty}^{\infty}\!\!d\omega\,f(\omega)e^{-i\omega t}=2\,\mathrm{Re}\int_0^{\infty}\!\!d\omega\,f(\omega)e^{-i\omega t}.
\end{equation}
Equation~\eqref{eq:EHGp} is written in terms of the electric-electric and magnetic-electric Green's function which are written as a sum of vacuum contributions $G^{(0)}$ and scattered parts $G^{\text{(sc)}}$, associated with the presence of the substrate. The former read
\begin{equation}\begin{split}
		\mathds{G}^{(0)}_\text{EE}(\omega,\mathbf{r},\mathbf{r}')&=\frac{e^{ik_0d}}{4\pi d}\Bigl[\Bigl(1+\frac{ik_0d-1}{k_0^2d^2}\Bigr)\mathds{I}\\
		&\,+\frac{3-3ik_0d-k_0^2d^2}{k_0^2d^2}\hat{\mathbf{d}}\otimes\hat{\mathbf{d}}\Bigr],\\
		\mathds{G}^{(0)}_\text{HE}(\omega,\mathbf{r},\mathbf{r}')&=\frac{e^{ik_0d}}{4\pi d}\frac{ik_0d-1}{k_0d^2}\begin{pmatrix}0 & -d_z & d_y\\d_z & 0 & -d_x\\-d_y & d_x & 0\end{pmatrix},
\end{split}\end{equation}
were $k_0=\omega/c$ and we have defined the distance vector $\mathbf{d} = \mathbf{r} - \mathbf{r}'$ having norm $d=|\mathbf{d}|$ and such that $\hat{\mathbf{d}}=\mathbf{d}/d$ and
\begin{equation}
	\mathbf{d} = d(\sin\theta_d\cos\varphi_d,\sin\theta_d\sin\varphi_d,\cos\theta_d).
\end{equation}
As discussed e.g. in Ref.~\cite{Novotny}, the scattering contribution to the Green's function, when both arguments $\mathbf{r}$ and $\mathbf{r}'$ are placed above the substrate ($z,z'>0$) can be written as
\begin{equation}\begin{split}
		\mathds{G}_\text{EE}&(\omega,\mathbf{r},\mathbf{r}')=\int\frac{dk}{2\pi}\frac{ike^{i\phi k_z(z+z')}}{2k_z}\Bigl[r_\text{TE}\begin{pmatrix}A & C & 0\\C & B & 0\\0 & 0 & 0\end{pmatrix}\\
		&+r_\text{TM}\frac{c^2}{\omega^2}\begin{pmatrix}-k_z^2B & k_z^2C & -\phi kk_zE\\k_z^2C & -k_z^2A & -\phi kk_z D\\\phi k k_zE & 
			\phi kk_z D& k^2F\end{pmatrix}\Bigr],\\
		\mathds{G}_\text{HE}&(\omega,\mathbf{r},\mathbf{r}')=\int\frac{dk}{2\pi}\frac{cke^{i\phi k_z(z+z')}}{2\omega k_z}\\
		&\times\Bigl[r_\text{TE}\begin{pmatrix}\phi k_z C & \phi k_z B & 0\\-\phi k_z A & -\phi k_z C & 0\\k D & - k E & 0\end{pmatrix}\\
		&-r_\text{TM}\begin{pmatrix}-\phi k_z C & \phi k_z A & k D\\-\phi k_z B & \phi k_z C & -k E\\0 & 0 & 0\end{pmatrix}\Bigr],
\end{split}\end{equation}
where
\begin{equation}
	\begin{pmatrix}A\\B\\C\\D\\E\\F\end{pmatrix}=\begin{pmatrix}\frac{1}{2}[J_0(kd) + J_2(kd)\cos(2\varphi_d)]\\ \frac{1}{2}[J_0(kd) - J_2(kd)\cos(2\varphi_d)]\\\frac{1}{2}J_2(kd)\sin(2\varphi_d)\\i J_1(kd)\sin\varphi_d\\ i J_1(kd)\cos\varphi_d\\J_0(kd)\end{pmatrix} 
	.\end{equation}
\begin{equation}
	\mathbf{p}_i(\omega) = \mathbf{p}_i^{\text{(fl)}}(\omega) + \frac{\omega^2}{c^2}\bar{\alpha}_i\sum_{j\neq i}\mathds{G}^{\text{(EE)}}(\mathbf{R}_i,\mathbf{R}_j)\mathbf{p}_j(\omega),
\end{equation}
where the field at the origin of each induced dipole is the total one except for the self contribution ($j\neq i$ in the sum) and we have introduced the polarizability $\bar{\alpha}_i$ of each dipole, which is a $3\times3$ matrix in the general scenario of an anisotropic particle. The previous equations lead to the linear system of equations
\begin{equation}
	\begin{pmatrix}\mathbf{p}_1(\omega)\\\vdots\\\mathbf{p}_N(\omega)\end{pmatrix} = \mathds{B}(\omega) \begin{pmatrix}\mathbf{p}^{\text{(fl)}}_1(\omega)\\\vdots\\\mathbf{p}^{\text{(fl)}}_N(\omega)\end{pmatrix},
\end{equation}
where $\mathds{B}(\omega) = \mathds{D}^{-1}(\omega)$ is a $3N\times3N$ matrix, $\mathds{D}$ being defined in terms of $3\times3$ blocks $\mathds{D}_{ij}$ ($i,j=1,\dots,N$) reading
\begin{equation}
\mathds{D}_{ij}(\omega) = \delta_{ij}\mathds{I} - (1 - \delta_{ij})\frac{\omega^2}{c^2}\bar{\alpha}_i\mathds{G}^{\text{(EE)}}(\omega,\mathbf{r}_i,\mathbf{r}_j).
\end{equation}
For the total electric and magnetic fields we have
\begin{equation}\begin{split}
		E_\alpha(\mathbf{r},\omega) &= \frac{\omega^2}{\varepsilon_0c^2}\mathds{G}^{\text{(EE)}}_{\alpha\beta}(\omega,\mathbf{r},\mathbf{r}_i) \mathds{B}_{ij,\beta\gamma}\,p^{\text{(fl)}}_{j,\gamma}(\omega),\\
		H_\alpha(\mathbf{r},\omega) &= -i\frac{\omega^2}{c}\mathds{G}^{\text{(HE)}}_{\alpha\beta}(\omega,\mathbf{r},\mathbf{r}_i)\mathds{B}_{ij,\beta\gamma}(\omega)p^{\text{(fl)}}_{j,\gamma}(\omega),
\end{split}\end{equation}
where from now on we use Latin indices for the index associated to the dipoles, going from 1 to $N$, and Greek ones for the Cartesian components $x$, $y$ and $z$, and sum over repeated indices is assumed.

\subsection{Poynting vector}

The component $\alpha$ of the Poynting vector can be easily put under the form
\begin{equation}
	S_\alpha(\mathbf{r}, t) = 2\,\mathrm{Re}\int_0^{+\infty}\!\frac{d\omega}{2\pi}e^{-i\omega t}S_\alpha(\mathbf{r},\omega),
\end{equation}
with
\begin{equation}S_\alpha(\mathbf{r},\omega)=\epsilon_{\alpha\beta\gamma}\int_0^{+\infty}\!\frac{d\omega'}{2\pi}\langle E_\beta(\mathbf{r},\omega')H_\gamma^*(\mathbf{r},\omega'-\omega)\rangle.
\end{equation}
where $\epsilon_{\alpha\beta\gamma}$ is the Levi-Civita tensor. We need at this stage the correlation function of the fluctuating dipoles, for which we employ the fluctuation-dissipation theorem
\begin{equation}
	\langle p^{\text{(fl)}}_{i,\alpha}(\omega)p^{\text{(fl)}*}_{j,\beta}(\omega') \rangle = 4\delta(\omega-\omega')\hbar\varepsilon_0\delta_{ij}\chi_{i,\alpha\beta}(\omega) n(\omega,T_i),\\
\end{equation}
depending on the individual dipole susceptibility matrix
\begin{equation}
	\bar{\chi}_i=\frac{\bar{\alpha}_i-\bar{\alpha}^\dag_i}{2i}-\frac{\omega^3}{6\pi c^3}\bar{\alpha}^\dag_i\bar{\alpha}_i,
\label{eq:chi}\end{equation}
and
\begin{equation}
 n(\omega,T) = \biggl[\exp\biggl(\frac{\hbar\omega}{k_B T}\biggr) - 1\biggr]^{-1}.
\label{eq:BE}\end{equation}
We deduce that the Poynting vector does not depend on time, as it should be for a stationary system, and reads
\begin{equation}
	S_\alpha(\mathbf{r}, t) = \int_0^{+\infty}\!\frac{d\omega}{2\pi}S_\alpha(\mathbf{r},\omega),
\end{equation}
the spectrum of the Poynting vector being given by
\begin{equation}\begin{split}
		S_\alpha&(\mathbf{r},\omega) = -\frac{4\hbar\omega^4}{c^3}\epsilon_{\alpha\beta\gamma}\\
		&\times\mathrm{Im}\Bigl[n(\omega,T_{i'})\mathds{G}^{\text{(EE)}}_{\beta\beta'}(\omega,\mathbf{r},\mathbf{r}_i)\mathds{G}^{\text{(HE)}*}_{\gamma\gamma'}(\omega,\mathbf{r},\mathbf{r}_j)\\
		&\qquad\times \mathds{B}_{ii',\beta'\beta''}(\omega)\mathds{B}^*_{ji',\gamma'\gamma''}(\omega)\,\chi_{i',\beta''\gamma''}(\omega)\Bigr].
\end{split}\label{eq:Sgeneral}\end{equation}

After obtaining the general expression given by Eq.~\eqref{eq:Sgeneral}, valid for an arbitrary number of dipole with anisotropic polarizabilities and in the presence of a substrate, we restrict ourselves to the scenario of a single dipole located at $\mathbf{r}_d$, having temperature $T_d$ and susceptibility $\bar{\chi}$. This scenario, as shown below, will already allow us to highlight effects of focusing of near-field radiative heat transfer ascribed to the anisotropic behavior. In this case we have $\mathds{B}=\mathds{I}$ and we deduce
\begin{equation}\begin{split}
		S_\alpha(\mathbf{r},\omega) &= -\frac{4\hbar\omega^4}{c^3}\epsilon_{\alpha\beta\gamma}\mathrm{Im}\Bigl[n(\omega,T_d)\mathds{G}^{\text{(EE)}}_{\beta\beta'}(\mathbf{r},\mathbf{r}_d)\\
		&\,\times\mathds{G}^{\text{(HE)}*}_{\gamma\gamma'}(\omega,\mathbf{r},\mathbf{r}_d)\,\chi_{\beta'\gamma'}\Bigr].
	\end{split}\label{eq:Sone}\end{equation}

\section{Results}\label{sec:results}

In this Section we are going to exploit Eq.~\eqref{eq:Sone} to compute the Poynting vector below a single particle in proximity to a substrate. As for the latter, we choose a substrate made of silicon carbide (SiC), whose permittivity is described here by a Drude-Lorentz model~\cite{PalikSiC}
\begin{equation}
	\varepsilon(\omega)=\varepsilon_\infty \frac{\omega^2_\mathrm{L}-\omega^2-i\Gamma\omega}{\omega^2_\mathrm{T}-\omega^2-i\Gamma\omega},
\end{equation}
with \textcolor{black}{high-frequency} dielectric constant $\varepsilon_\infty=6.7$, longitudinal optical frequency $\omega_\mathrm{L}=1.83\times 10^{14}\,$rad/s, transverse optical frequency $\omega_\mathrm{T}=1.49\times 10^{14}\,$rad/s, and damping $\Gamma=8.97\times 10^{11}\,$rad/s. According to this model a SiC substrate supports a surface phonon-polariton mode in $p$ polarization at frequency $\omega_\text{pl} = 1.786\times 10^{14}\,$rad/s (solution in the large-wavevector limit of the equation $\varepsilon(\omega) + 1 = 0$), which is expected to contribute significantly to near-field effects such as the one studied in this work.

\begin{figure}[t!]
	\includegraphics[width=0.99\linewidth]{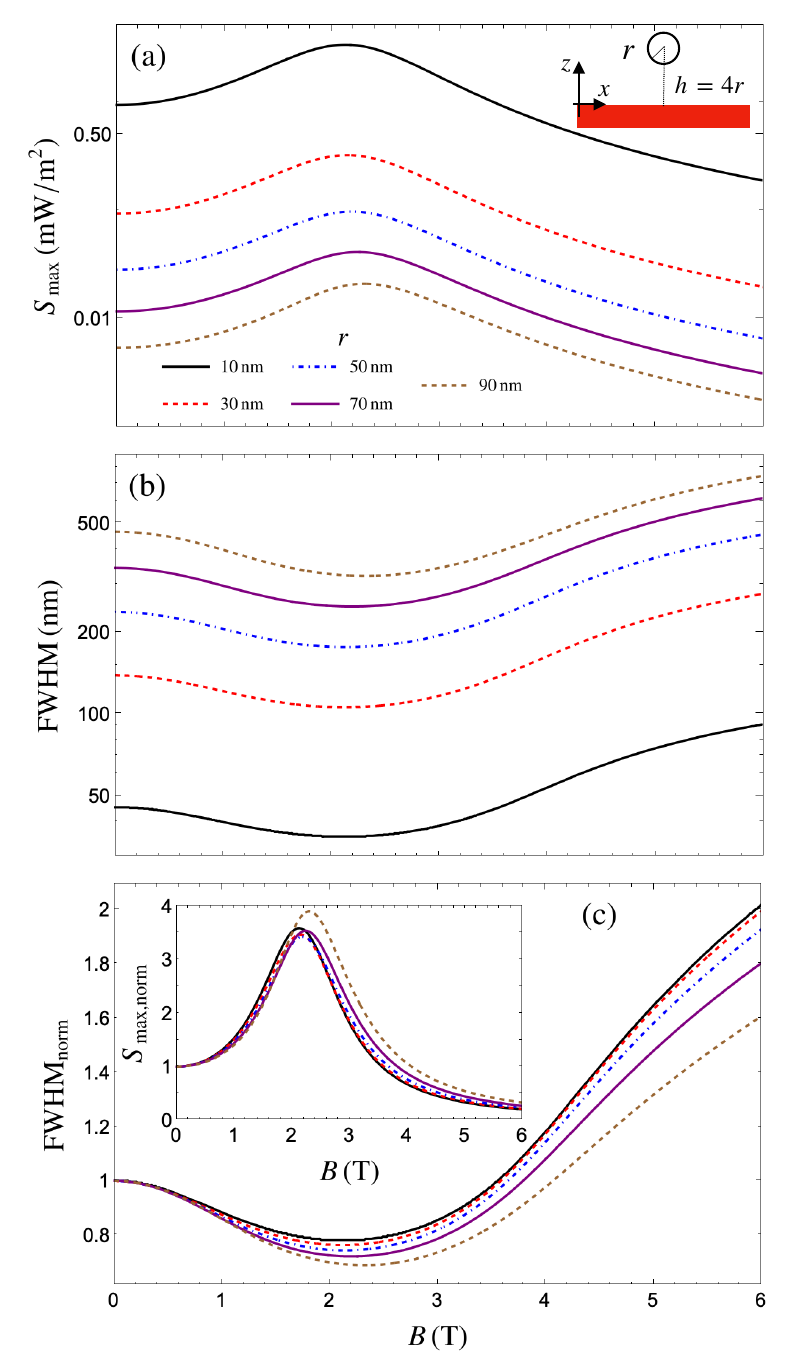}
	\caption{(a) Poynting vector $S_\mathrm{max}$ below the nanoparticle, placed at $(0,0,4r)$, and (b) FWHM as a function of the applied magnetic field $B$ for different radii (see legend). Panel (c) and its inset show the same curves normalized by the value at $B=0$.}
	\label{fig:SFBr}
\end{figure}

Concerning the nanoparticle, we assume that it is made of indium antimonide (InSb), a magneto-optical material whose properties have been recently studied in connection with their impact on near-field radiative heat transfer. This material is characterized by an isotropic permittivity, becoming anisotropic in the presence of a magnetic field, allowing us to address the impact of this induced anisotropy on the Poynting vector. More specifically, in the presence of a magnetic field $\mathbf{B}=B\hat{\mathbf{e}}_z$ acting along the $z$ direction, the permittivity matrix $\bar{\varepsilon}$ takes the form
\begin{equation}
	\bar{\varepsilon}=\begin{pmatrix} \varepsilon_1 & -i\varepsilon_2 & 0 \\ i\varepsilon_2 & \varepsilon_1 & 0 \\ 0 & 0 & \varepsilon_3 \end{pmatrix},
\label{eq:permInSb}\end{equation}
where
\begin{equation}
	\begin{split}
		\epsilon_1 &= \epsilon_\infty\biggl(1+\frac{\omega_{\rm L}^2-\omega_{\rm T}^2}{\omega_{\rm T}^2-\omega^2-i\Gamma\omega} +\frac{\omega_{\rm p}^2(\omega+i\gamma)}{\omega[\omega_{\rm c}^2-(\omega+i\gamma)^2]} \biggr), \\
		\epsilon_2 &= \frac{\epsilon_\infty\omega_{\rm p}^2\omega_{\rm c}}{\omega[(\omega+i\gamma)^2-\omega_{\rm c}^2]},\\ 
		\epsilon_3 &= \epsilon_\infty\left(1+\frac{\omega_{\rm L}^2-\omega_{\rm T}^2}{\omega_{\rm T}^2-\omega^2-i\Gamma\omega}-\frac{\omega_{\rm p}^2}{\omega(\omega+i\gamma)} \right),
	\end{split}
\end{equation}
$\omega_c = e B /m^*$ being the cyclotron frequency and where the parameters read~\cite{Ben-Abdallah2016} $\varepsilon_\infty=15.7$, $\omega_\mathrm{L}=3.62\times 10^{13}\,$rad/s, $\omega_\mathrm{T}=3.39\times 10^{13}\,$rad/s, $\Gamma=5.65\times 10^{11}\,$rad/s, $n = 1.36\times10^{19}$\,cm$^{-3}$, $m^* = 7.29\times10^{-32}$\,kg, $\omega_{\rm p} = \sqrt{\frac{ne^2}{m^*\epsilon_0\epsilon_\infty}} = 1.86\times10^{14}$\,rad/s, and $\gamma = 10^{12}$\,rad/s. The permittivity leads to the polarizability tensor $\bar{\alpha}$ defined as~\cite{LakhtakiaEtAl1991,Albaladejo}
\begin{equation}
	\bar{\alpha} = 4\pi R^3(\bar{\varepsilon}-\mathds{I})(\bar{\varepsilon}+2\mathds{I})^{-1},
\end{equation}
which finally allows to deduce the susceptibility through Eq.~\eqref{eq:chi}. It is worth mentioning that in the case of nanoparticles described within the dipolar approximation near-field effects strongly depend on dipolar resonances (existing at the interface between each nanoparticle and the surrounding vacuum). In the case of a permittivity given by Eq.~\eqref{eq:permInSb} and in the limit of absence of dissipation these resonances can be deduced analytically and are located at~\cite{Ott2019b}
\begin{equation}
	\begin{split}
		\omega_{m = \mp1} &= \sqrt{\left(\frac{\epsilon_\infty\omega_{\rm p}^2}{\epsilon_\infty+2}+\frac{\omega_{\rm c}^2}{4}\right)} \pm \frac{\omega_{\rm c}}{2}, \\
		\omega_{m = 0} &= \sqrt{\frac{\epsilon_\infty\omega_{\rm p}^2}{\epsilon_\infty+2}}.
\end{split}\label{eq:resInSb}\end{equation}
We clearly see that these are degenerate for $B = 0$, while for $B\neq 0$ the resonances with $m=\mp1$ deviate approximately linearly from the one having $m=0$, with a slope proportional to the cyclotron frequency $\omega_{\rm c}$.

\begin{figure}[t!]
	\includegraphics[width=0.99\linewidth]{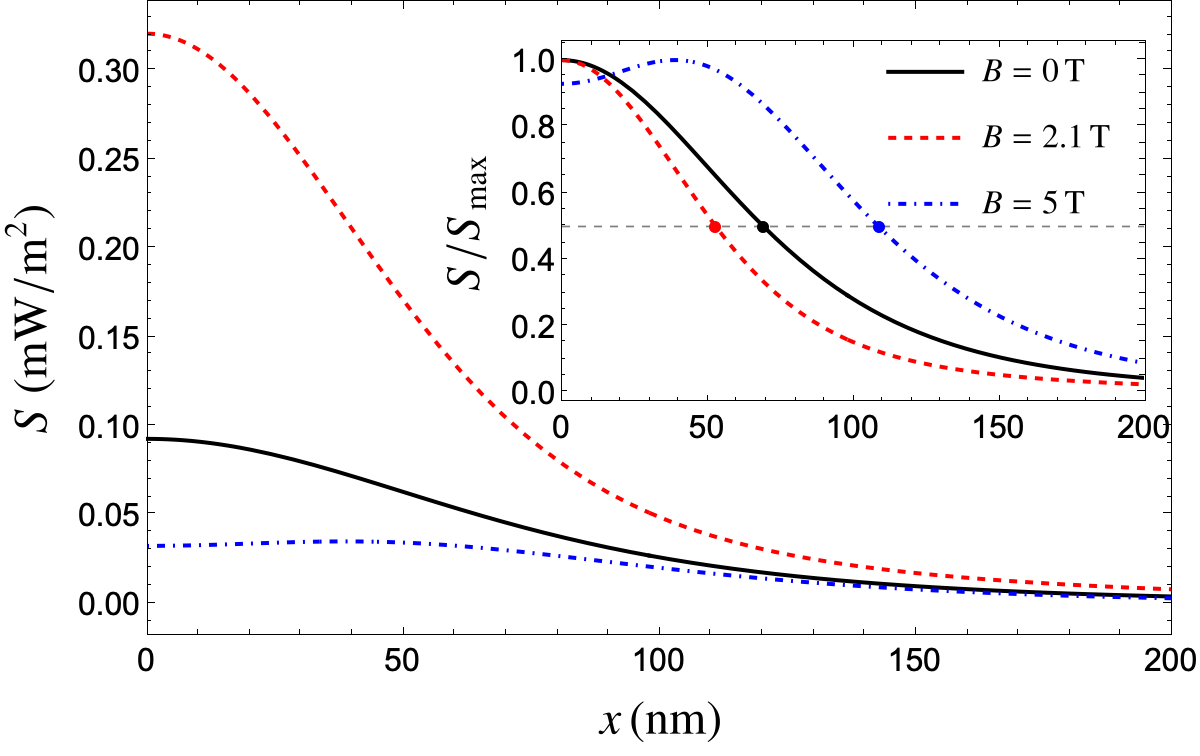}
	\caption{Poynting vector along the $x$ axis for three values of the applied field (see legend), in absolute and normalized (inset) units. In the inset, the points indicate the coordinate where FWHM is realized.}
	\label{fig:Sx}
\end{figure}

By using these definition, we are going to focus in the following on two main quantities to assess the performance of the system in terms of HAMR. The first is the $z$ component of the Poynting vector $S_z(0,0,0)$ at the location right below the nanoparticle and at the interface between substrate and vacuum. This quantity, that we will define as $S_\mathrm{max}$, corresponds (as shown below) to the highest value of $S_z$ on the plane $z=0$ for vanishing and moderate values of the magnetic field, and a potential increase due to the magnetic field anticipates the possibility of increasing heat transfer between nanoparticle and substrate and thus making a local temperature increase of the substrate easier. The second relevant quantity is the Full Width Half Maximum (FWHM), defined as the $2x_0$, $x_0$ being the coordinate at which $S_z(x_0,0,0)=S_z(0,0,0)/2$. This quantity represents a measure of the localized nature of the heat transfer between nanoparticle and substrate and thus of the potential spatial resolution of the associated HAMR device.

\begin{figure}[t!]
	\includegraphics[width=0.86\linewidth]{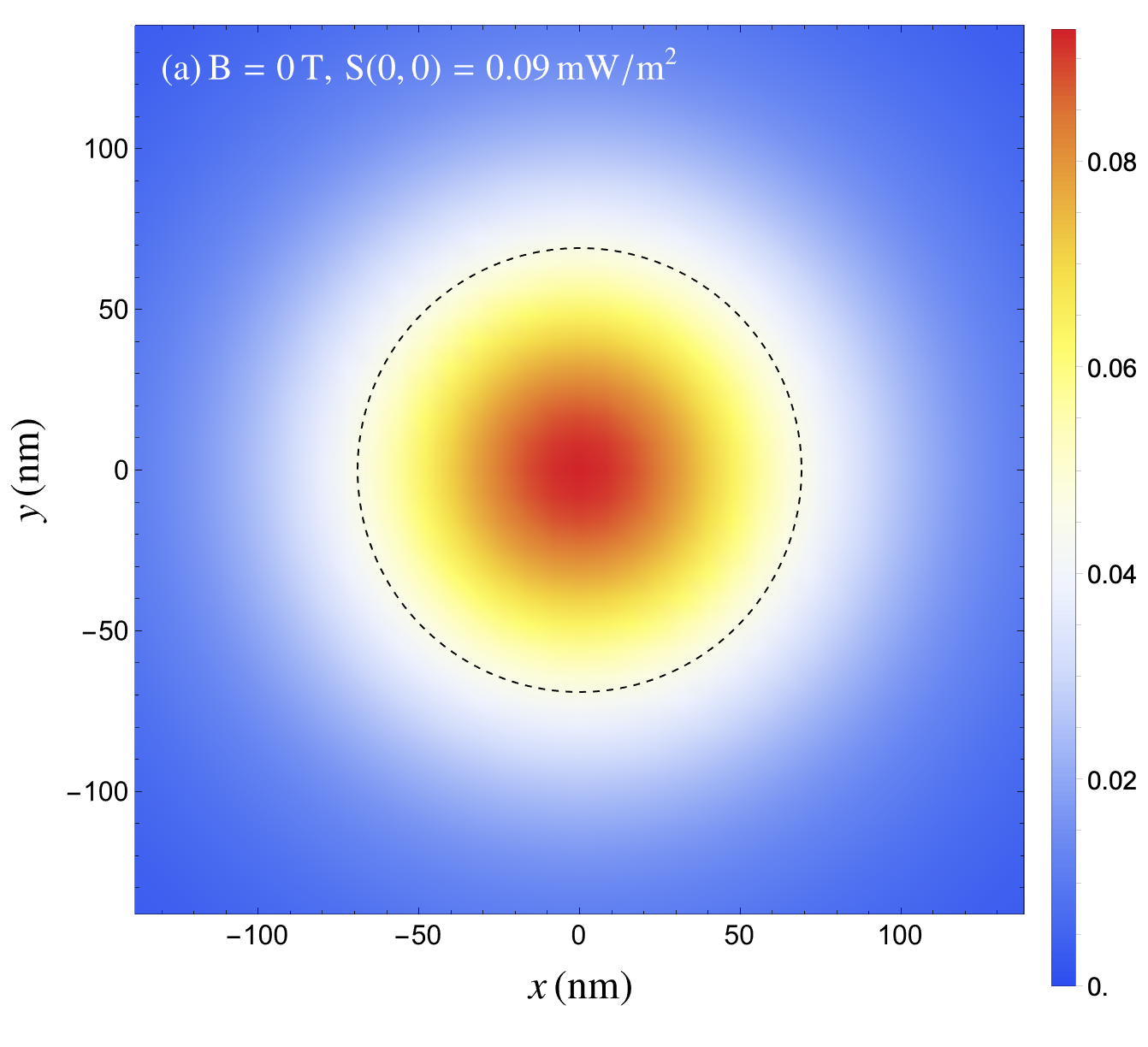}
	\includegraphics[width=0.86\linewidth]{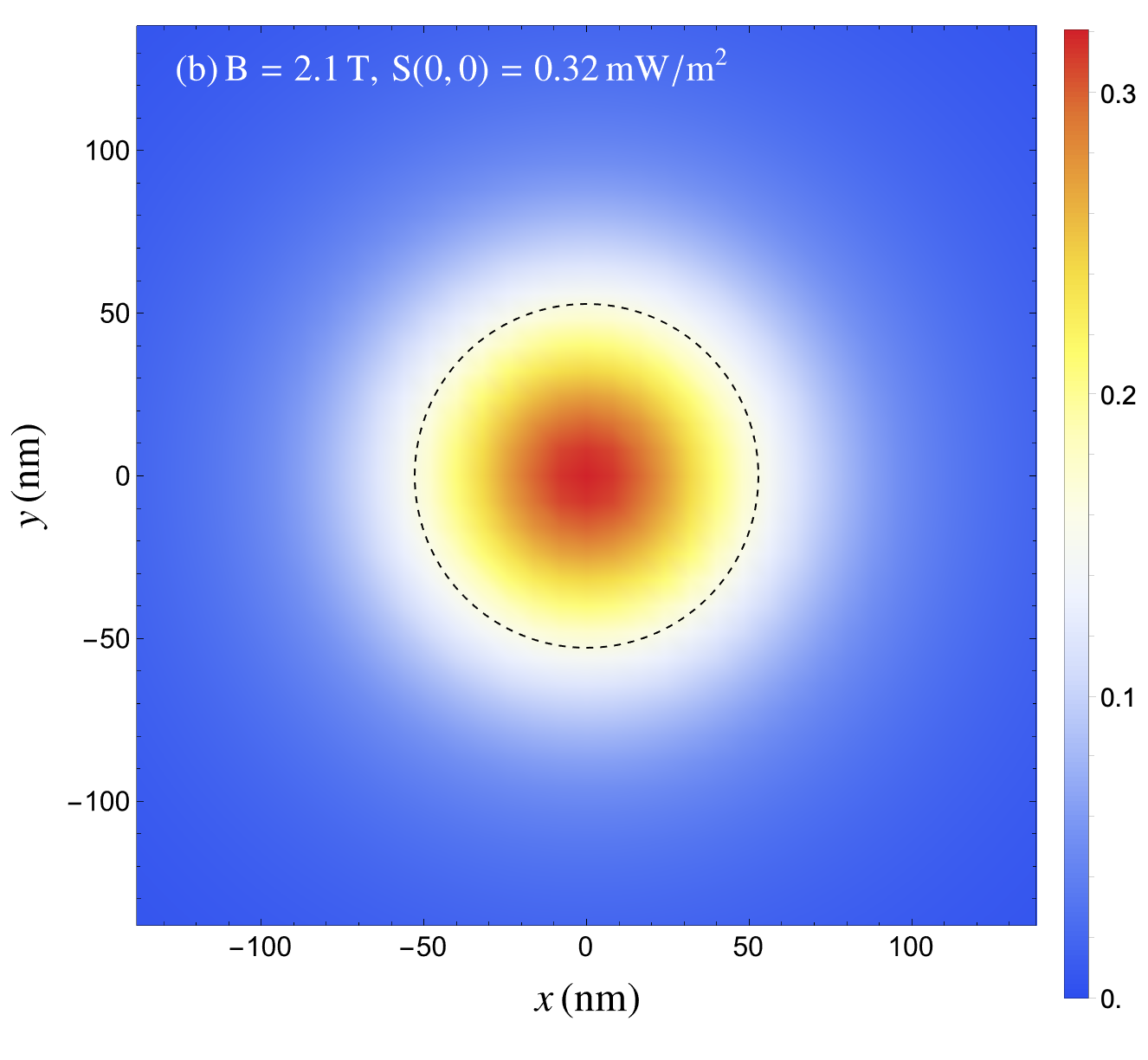}
	\includegraphics[width=0.86\linewidth]{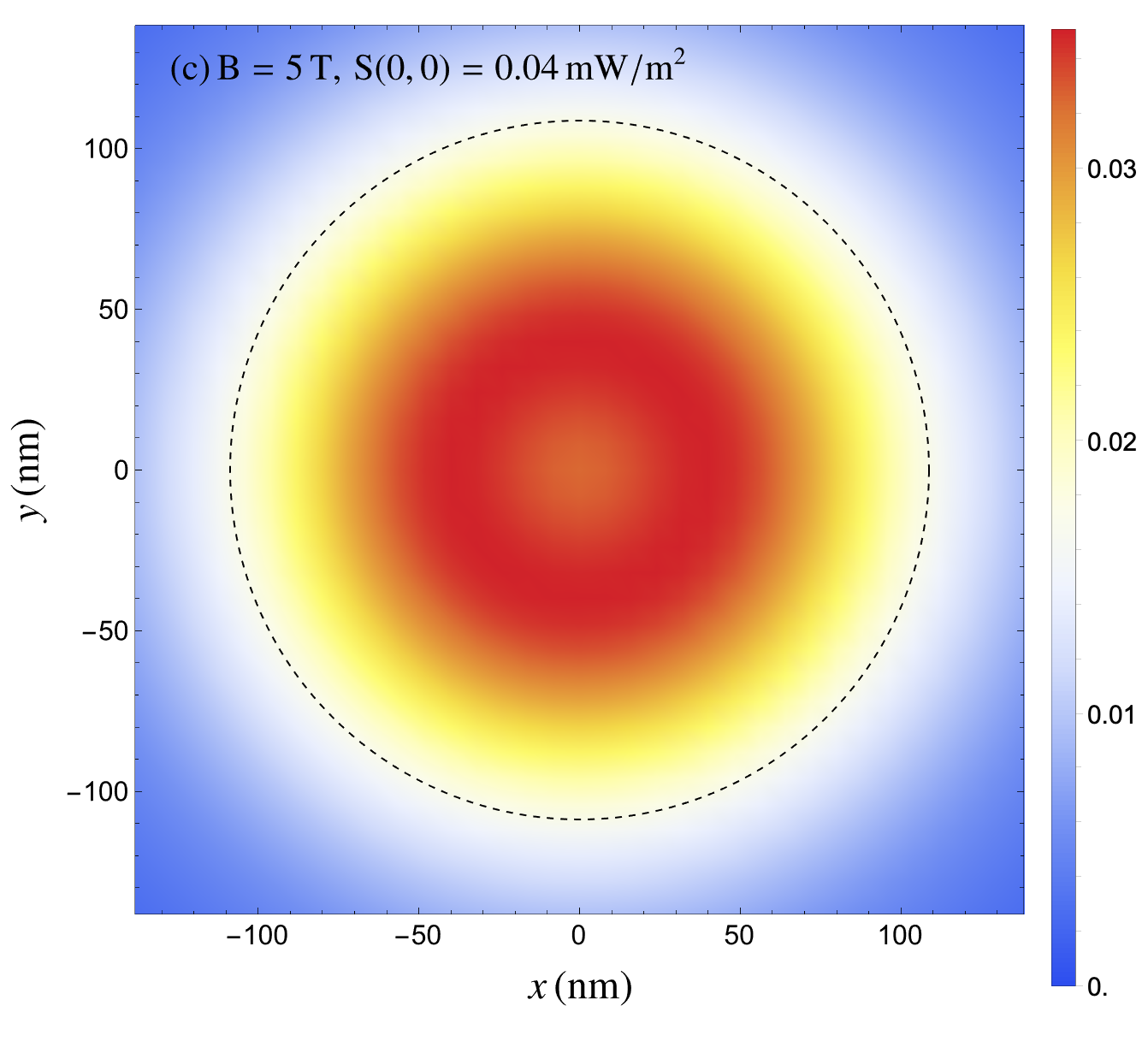}
	\caption{$z$ component of the Poynting vector $S_z(x,y,0)$ on the $z$ plane for different values of the applied magnetic field (see legend). In each panel, the dashed circle represents the width (FWHM) of the Poynting vector at the surface of the substrate.}
	\label{fig:Sxy}
\end{figure}

As a first result, we study $S_\mathrm{max}$ and FWHM as a function of the applied magnetic field $B$ for different radii $r$ of the nanoparticle. For each $r$, in order to guarantee the validity of the dipolar approximation, we choose the coordinates of the particle as $(0,0,h)$ with $h=4R$, as shown in the inset of Fig.~\ref{fig:SFBr}(a). Figures \ref{fig:SFBr}(a)-(b) show the results for radii $r=10,30,50,70,90\,$nm (and thus distances $h=40,120,200,280,360\,$nm from the substrate) as a function of $B$ in a range from 0 to $6\,$T. Let us first focus on the isotropic scenario $B=0$ corresponding to the absence of applied magnetic field. Concerning $S_\mathrm{max}$, we remark that its value increases when decreasing the radius $r$. While this might seem surprising because of the reduced particle polarizability, it is a signature of the strong near-field dependence of $S_\mathrm{max}$ on the distance, which decreases since proportional to the radius. Concerning the FWHM, we observe that it increases with the radius (thus with the distance $h$), and that is always of the order of $h$.

Let us now focus on the impact of the magnetic field. First, we remark a non-monotonic behavior, according to which $S_\mathrm{max}$ (FWHM) first increases (decreases) as a function of $B$, then decreases (increases), by reaching values even below (above) the reference one for $B=0$. This clearly highlights an ideal scenario as a function of $B$, since it shows the existence of an optimal magnetic field $B_\mathrm{opt}$ for which not only is the local heat flux quantitatively increased, but it is also more focused on the surface of the substrate. The relative increase (decrease) of $S_\mathrm{max}$ (FWHM) is shown in Fig.~\ref{fig:SFBr}(c), where we observe that the relative effect does not depend strongly on the particule radius $r$, with $S_\mathrm{max}$ increasing almost by a factor of 4 and FWHM being reduced by around 30\% in the best scenario. We remark that the two phenomena take place around the same optimal value for $B$, which is here $B_\mathrm{opt}\simeq2.1\,$T.

\begin{figure}[t!]
	\includegraphics[width=0.99\linewidth]{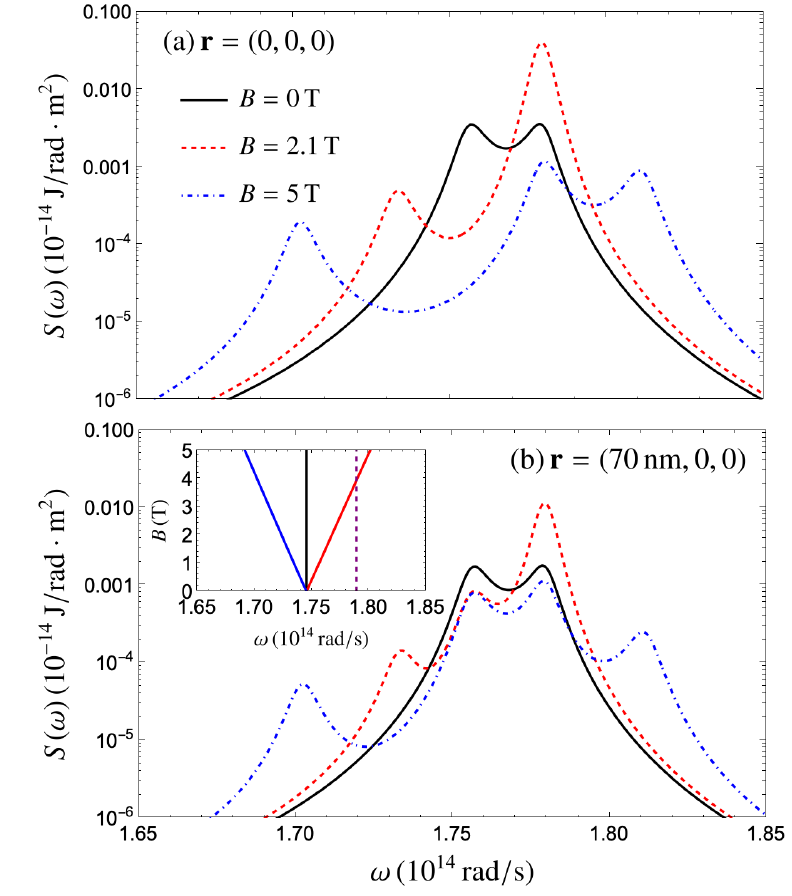}
	\caption{Spectral decomposition of the $z$ component of the Poynting vector at (a) the origin and (b) at coordinate $(70\,\mathrm{nm},0,0)$, for different values of the applied magnetic field (see legend). The inset of panel (b) represents the three resonances of InSb defined in Eq.~\eqref{eq:resInSb} (black for $m=0$, red for $m=-1$ and blue for $m=+1$), while the vertical dashed purple line represents the resonance frequency of SiC.}
	\label{fig:Somega}
\end{figure}

Before addressing more in detail the origin of the optimal field, it is interesting to analyze the shape of $S(x,0,0)$ along the $x$ axis for three values of the magnetic field $B=0,2.1,5\,$T, namely a vanishing field, the optimal one, and a field for which the result is worse (in terms of HAMS application) than the one in the absence of applied field. This is shown in Fig.~\ref{fig:Sx}, both in absolute units and in normalized ones (inset). As expected, we remark an almost 4-fold enhancement of the Poynting vector right below the nanoparticle, and the inset in normalized units allows us to clearly visualize the width reduction.

A complementary view of the impact of the magnetic field on $S_\mathrm{max}$ and FWHM is given by Fig.~\ref{fig:Sxy}, where for the same values of the magnetic field $S_z(x,y,0)$ is shown on the plane $z=0$.

In order to get more insight on the existence of an optimal field $B_\mathrm{opt}=2.1\,$T, we now perform a more detailed spectral investigation of the effect. We first analyze the spectrum $S(\omega)$ of the Poynting vector at $\mathbf{r}=(0,0,0)$, i.e. right below the nanoparticle at the location of $S_\mathrm{max}$, shown in Fig.~\ref{fig:Somega}. For clarity, let us start from the spectrum corresponding to $B=0$ [black solid curve in Fig.~\ref{fig:Somega}(a)]. We remark the presence of two resonances. The one at higher frequencies corresponds to the resonance at a SiC--vacuum interface discussed above. On the contrary, the one at lower frequency stems from the InSb nanoparticle resonances, all degenerate for $B=0$.

As discussed above, the presence of a magnetic field induces the appearance of two further resonances, whose deviation from the one at $B=0$ scales almost linearly with $\omega_{\rm c}$, thus with $B$. The effect is manifest in Fig.~\ref{fig:Somega}(a), where we remark that $B_\mathrm{opt}=2.1\,T$ is the value of the field for which one of this resonances matches the one of SiC (independent of the applied field $B$), thus leading to a strong amplification of $S_\mathrm{max}$. The possibility of obtaining this matching condition with a relatively moderate magnetic field is indeed connected to the proximity of SiC and InSb resonance in the absence of magnetic field. In the inset of Fig.~\ref{fig:Somega}(b), we show the three field-dependent resonance frequencies of InSb defined in Eq.~\eqref{eq:resInSb} as a function of the applied field $B$, along with the resonance frequency of SiC. While this curve is useful to visualize the appearance of a matching as a function of $B$, it does not allow to quantitatively extract the corresponding optimal value of the magnetic field because both the resonant frequencies given in Eq.~\eqref{eq:resInSb} and the one of SiC are deduced by assuming the absence of losses. Going back to Fig.~\ref{fig:Somega}(a), when increasing the field $B$ further above $B_\mathrm{opt}$, not only does the frequency matching disappear, but the resonance frequencies associated with InSb approach outer regions of the spectrum, and as a consequence induce a smaller heat transfer because of the Planck window dictated by the temperature mainly through the Bose-Einstein distribution \eqref{eq:BE}.

While this frequency-matching effect explains the amplification of local heat transfer right below the nanoparticle, i.e. at the origin, it does not account for the reduction of FWHM. In order to explain this further relevant effect of anisotropy, we present in Fig.~\ref{fig:Somega}(b) the same spectral analysis as before but at the different coordinate $(70\,\mathrm{nm},0,0)$, corresponding to half the FWHM realized at $B=0$. We observe the same qualitative behavior discussed above, i.e. a maximum enhancement occurring at $B_\mathrm{opt}=2.1\,$T because of a frequency-matching condition. Nevertheless, the enhancement at the peak frequency (the resonance of SiC), giving the largest contribution to the frequency-integrated Poynting vector, is less pronounced. This can be attributed to the fact that the almost monochromatic nature of the heat flux is a purely near-field effect, depending strongly on the distance. As a consequence, it is more pronounced right below the nanoparticle (at a distance $h=120\,$nm) than at coordinate $(70\,\mathrm{nm},0,0)$ (at a distance $\simeq140\,$nm).The result is that, although the heat flux is amplified on the entire surface $z=0$, the distance-dependent amplification is stronger right below the nanoparticle and is at the origin of the desired reduction of FWHM.

\section{Conclusion}

We have introduced a theoretical framework to calculate the Poynting field radiated by a system of $N$ magneto-optical thermal emitters placed above an isotropic substrate within the dipolar approximation. By applying this formalism to the specific scenario of a single particle we have shown that the application of an external magnetic field can induce both an amplification of the flux of Poynting vector at the substrate interface and a substantial decrease in the heated zone's surface area. This result paves the way to significant performance improvements for the HAMR technology.

\end{document}